\documentclass[11pt,a4paper]{article}

\usepackage[T1]{fontenc}
\usepackage[utf8]{inputenc}
\usepackage{lmodern}
\usepackage[a4paper,margin=2.5cm]{geometry}
\usepackage[authoryear,round]{natbib}
\usepackage{url,lineno,microtype}
\usepackage[onehalfspacing]{setspace}
\usepackage{amsmath,amssymb,amsfonts}
\usepackage{graphicx}
\usepackage{multirow}
\usepackage{booktabs}
\usepackage{array,ragged2e}
\usepackage{braket}
\usepackage[caption=false,font=footnotesize]{subfig}
\usepackage{hyperref}
\graphicspath{{./}{./assets/}}

\newcommand{\tabstyle}{\small\setlength{\tabcolsep}{4pt}}

\title{Quantum Recurrent Architectures for Financial Time Series Prediction}

\author{Danyal Maheshwari$^{1,*}$, Gerhard Hellstern$^{1}$,
Martin Zaefferer$^{2}$, Martin Braun$^{3}$ and Tanja D\"ohler$^{3}$\\[1.5ex]
\small $^{1}$Zentrum f\"ur Finanzwirtschaft, DHBW Stuttgart, Stuttgart, Germany\\
\small $^{2}$Zentrum f\"ur Digitale Innovationen, DHBW Ravensburg, Ravensburg, Germany\\
\small $^{3}$DATEV eG, N\"urnberg, Germany\\[1ex]
\small $^{*}$Correspondence: Danyal Maheshwari, \href{mailto:danyal.maheshwari@dhbw-stuttgart.de}{danyal.maheshwari@dhbw-stuttgart.de}}

\date{}

\begin{document}

\maketitle

\begin{abstract}
This study evaluates quantum classical hybrid recurrent architectures for financial time-series forecasting: a Quantum Long Short-Term Memory (QLSTM) network and Quantum Reservoir Computing (QRC) with lag-based inputs. Classical lag vectors are encoded into quantum states via amplitude encoding, which compactly represents normalized observations but requires nontrivial state preparation; the QLSTM gates and the QRC reservoir are implemented with parameterized quantum circuits, while classical optimizers train the interface and readout layers. On a monthly revenue series extended to 1056 months by a Gaussian-process--hidden-Markov generator, with a fixed chronological 70/15/15 split and one-step-ahead forecasts, we compare both models against a full-size LSTM, a classical control matched in parameter budget and interface architecture, a small plain LSTM, and classical reservoir computing, over ten seeds and against persistence, seasonal-naive and SARIMA benchmark forecasts. The QLSTM, with 2,105 parameters, is statistically indistinguishable from an LSTM 126 times larger, but equally indistinguishable from the matched classical control and from a plain LSTM of the same budget, so its parameter efficiency cannot be attributed to the quantum circuits. The quantum reservoir is worse than a classical reservoir of equal feature dimension in all ten seeds and worse than SARIMA, whereas all recurrent models and both classical reservoirs improve on SARIMA by about 2\% in RMSE. In a 24-month rolling-origin backtest the classical LSTMs degrade more slowly than the QLSTM. We find no evidence of a quantum accuracy or parameter-efficiency advantage in this low-dimensional setting.

\end{abstract}

\section{Introduction}
Forecasting business revenue series is difficult: the histories are short, the series combine trend, seasonality and abrupt level shifts, and simple statistical benchmarks are often competitive with machine-learning forecasters \citep{makridakis2018}. Financial series more broadly are documented to be non-stationary and heavy-tailed and to be subject to regime shifts and structural breaks \citep{schmitt2013,cont2001,lux2009}. Recurrent neural networks, and in particular Long Short-Term Memory (LSTM) networks \citep{hochreiter1997}, are standard tools for such data \citep{fischer2018}, but they carry large parameter counts. Reservoir computing (RC) \citep{jaeger2001} is a lightweight alternative in which a fixed random recurrent system generates nonlinear features and only a linear readout is trained. Quantum machine learning \citep{biamonte2017,schuld2015,schuld2018} has motivated quantum counterparts of both: the Quantum LSTM (QLSTM) \citep{chen2022qlstm} replaces the affine gate maps of an LSTM cell by variational quantum circuits (VQCs), and quantum reservoir computing (QRC) \citep{fujii2017,ghosh2019,mujal2021} uses a fixed quantum system as the reservoir. Both have been applied to financial problems \citep{doosti2024,egger2020,orus2019}.

\citet{chen2022qlstm} introduced the QLSTM and evaluated it on synthetic signals (damped oscillations, Bessel functions, and similar) against a classical LSTM, reporting faster convergence but no parameter-matched control. Several follow-up studies apply QLSTM to real series: \citet{khan2023qlstm} compare QLSTM and LSTM on solar-power forecasting; \citet{cao2023qlstm} add a linear layer to the QLSTM for carbon-price forecasting; \citet{mahmood2024qlstm} compare LSTM and QLSTM for stock-market movement; \citet{ceschini2022} study hybrid quantum--classical RNNs for time-series prediction. In each of these the classical baseline is a full-size LSTM, and the reported experiments use one or a few runs; none includes a classical model matched to the hybrid's parameter budget \emph{and} interface architecture, and none reports seasonal-naive or ARIMA-type benchmark forecasts.

On the reservoir side, \citet{fujii2017} and \citet{ghosh2019} established QRC; \citet{chen2020nurdin} and \citet{fry2023} apply gate-based and noise-induced QRC to nonlinear and chaotic benchmarks; \citet{mujal2021} review QRC and quantum extreme learning machines, covering computational principles, physical implementations and applications. Financial applications include realized-volatility forecasting \citep{li2026qrc} and credit-default classification on neutral-atom hardware \citep{vitali2025}. These works compare QRC to echo-state networks or to classical classifiers, but again not with repeated seeds against forecasting benchmarks on real revenue-type data.

Finally, \citet{bowles2024} showed in a large classification benchmark that out-of-the-box classical models systematically outperform small quantum models and that removing entanglement often does not hurt; \citet{makridakis2018} showed that on real business time series, statistical benchmarks are frequently competitive with machine-learning forecasters.

Hybrid quantum recurrent models are motivated in part by the possibility of achieving useful forecasting performance with fewer trainable parameters. A comparison with a full-size classical network alone cannot determine whether performance reflects the quantum gates, the smaller parameter budget, or the model architecture. We therefore evaluate QLSTM against a full-size LSTM and cLSTM-S, a classical model matched in parameter budget and interface architecture. The QLSTM has 2,105 trainable parameters, including 48 quantum parameters, compared with 265,473 for the full-size LSTM. We also compare QRC with a classical reservoir with the same feature dimension. All models use matched data splits and training protocols and are evaluated against persistence, seasonal-naive, and SARIMA baselines.

This paper contributes the following:
\begin{enumerate}
\item A common forecasting protocol for QLSTM, QRC, LSTM and RC on a monthly revenue series extended to 1056 months by a GP--HMM generator, with a fixed chronological 70/15/15 split and one-step-ahead forecasts (Section~\ref{sec:data}).
\item Matched classical controls: cLSTM-S, identical to the QLSTM except that each 2-qubit VQC is replaced by a classical block of the same input/output dimension; LSTM-S, a plain LSTM of the same parameter budget; and a classical reservoir with the same feature dimension and readouts as the QRC.
\item Evaluation over 10 seeds against three baseline forecasts with MSE, RMSE, MAE, and MASE; paired Wilcoxon tests with Holm correction; a training-window ablation; a 24-month rolling-origin backtest; and wall-clock cost (Section~\ref{sec:results}).
\end{enumerate}

The remainder of this paper is organized as follows. Section~\ref{sec:methods} describes the financial dataset, lag construction and preprocessing, the classical and quantum models, the benchmark forecasts and the experimental protocol. Section~\ref{sec:results} presents and discusses the forecasting results, including the limitations of the study, and Section~\ref{sec:conclusion} concludes.

\section{Materials and Methods}\label{sec:methods}

\subsection{Financial data and protocol}\label{sec:data}

\subsubsection{Raw data}
This work investigates time-series forecasting on a financial dataset. We use a single univariate monthly revenue series of one product, observed for eight years (96 months).

\begin{table}[h!]
\centering
\caption{GP kernel components, initial values and parameter limits ($\ell$ are length scales and $p$ is the period, both in years).}
\label{tab:kernels}
\renewcommand{\arraystretch}{1.1}
\small
\setlength{\tabcolsep}{3pt}
\begin{tabular}{p{0.19\linewidth}p{0.25\linewidth}p{0.2\linewidth}p{0.28\linewidth}}
\toprule
\textbf{Component} & \textbf{Purpose} & \textbf{Initialization} & \textbf{Constraint / limit} \\
\midrule
Rational quadratic & Slow trend and medium-scale variation & $\ell_{\text{RQ}}=4.0$, $\alpha=0.5$ & $\ell_{\text{RQ}}\in[0.8,10.0]$ \\
Mat\'ern (short-term) & Short-term variability & $\ell_{\text{short}}=0.4$ & $\ell_{\text{short}}\in[0.1,0.7]$, fixed $\nu=1.5$ \\
Mat\'ern (long-term) & Smooth medium- to long-term variation & $\ell_{\text{long}}=1.0$ & $\ell_{\text{long}}\in[0.7,3.0]$, fixed $\nu=2.5$ \\
Periodic & Seasonality & $\ell_{\text{Per}}=1.0$ & $\ell_{\text{Per}}\in[0.1,2.0]$, $p\in[0.1,1.5]$ \\
\bottomrule
\end{tabular}
\end{table}

\subsubsection{Synthetic data generation}
A history of 96 observations is comparatively short for studying long-range forecasting behavior, so we extend it with synthetic continuations that preserve important characteristics of the raw data. Gaussian processes (GPs) are well suited for this purpose because they can represent fast changes, slower trends, and periodic behavior. However, many revenue series show a slowly varying mean together with episodes in which the values stay for some time at a lower or higher level, and a plain GP represents such behavior only imperfectly. We therefore augment the GP with a two-state hidden Markov model (HMM) on the residual process, so that synthetic samples combine smooth GP dynamics with state-dependent local level shifts.

Only the first 60 months of the series are used to fit the GP--HMM; the remaining 36 months are not used by the generator. The raw observations $y_t$ are transformed by clipping negative values, applying a $\log(1+y)$ transform, and standardizing with the mean $\bar{u}$ and standard deviation $s$ of the transformed training months:
\begin{align}
\tilde{y}_{t} &= \max(y_{t},0), &
u_{t} &= \log(1+\tilde{y}_{t}), &
x_{t} &= \frac{u_{t}-\bar{u}}{s}.
\end{align}
Time is rescaled from months to years via $\tau_t = (m_t-\min_t m_t)/12$, where $m_t$ is the month index.

The latent smooth component of the series is modeled by an exact Gaussian process
\begin{equation}
f(\tau) \sim \mathcal{GP}\!\left(\mu(\tau), k(\tau,\tau')\right),
\end{equation}
with a constant mean and an additive kernel combining a rational quadratic component, two Mat\'ern components and a periodic component, each wrapped in a scale kernel. The components and their hyperparameters are listed in Table~\ref{tab:kernels}. This follows the standard GP regression framework of \citet{rasmussenwilliams2006} and is implemented with GPyTorch \citep{gardner2018}.

The HMM models the residual of the GP, using standard forward--backward and Baum--Welch estimation \citep{rabiner1989,baum1970}. The resulting observation model is
\begin{equation}
x_t = f(\tau_t) + o_t + \varepsilon_t,
\qquad
\varepsilon_t \sim \mathcal{N}(0,\sigma_\varepsilon^2),
\end{equation}
where $s_t\in\{1,2\}$ is the hidden state and $o_t=o_{s_t}$ is the state-dependent offset.

The fitting procedure alternates between the GP and HMM components across five outer rounds. In each round, the GP is optimized for 500 gradient steps, and the HMM is then fit to the residuals for up to 50 Baum--Welch iterations. To reduce sensitivity to the initialization of the periodic kernel, the entire procedure is repeated using six different initial values for the period parameter; the final model is then selected based on the lowest negative marginal log-likelihood of the GP training data.

Synthetic data are obtained by evaluating the trained GP on an extended query grid, drawing GP sample paths, sampling HMM state trajectories from the learned Markov chain, and adding Gaussian observation noise. The samples are then mapped back to the original scale using
\begin{equation}
\hat{y}_{t} = \exp\!\left(s\, \hat{x}_{t} + \bar{u}\right)-1.
\end{equation}
This procedure yields synthetic series that preserve smooth trend and seasonal structure as well as local level shifts.

\subsubsection{Experimental dataset and split}
\begin{figure}[h!]
\centering
\includegraphics[width=\textwidth]{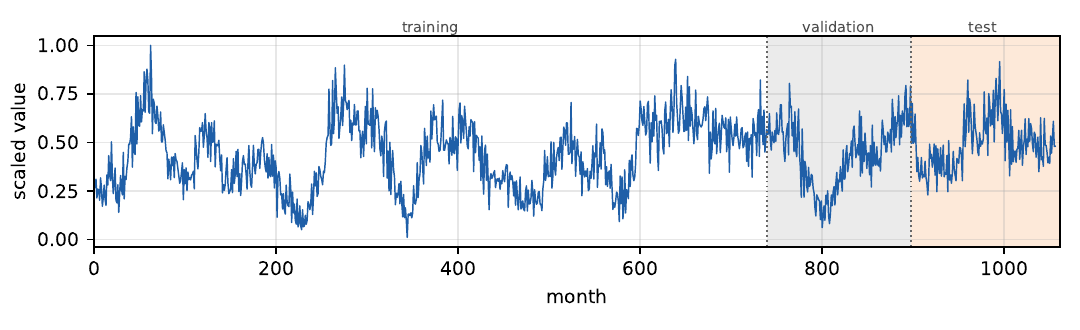}
\caption{The experimental series (1056 months, min--max scaled on the training window), coloured by the chronological 70/15/15 split into training (months 1--739), validation (740--897) and test (898--1056) windows.}
\label{fig:synth}
\end{figure}

The query grid covers 1056 months: the observation period followed by a 960-month continuation, so the resulting series contains 1056 monthly values drawn from the fitted GP--HMM (Figure~\ref{fig:synth}). It is split chronologically into a training window (months 1--739, 70\%), a validation window (months 740--897, 15\%), and a test window (months 898--1056, 15\%, corresponding to 159 months). The validation window is used only for early stopping and readout selection; the test window is used once, for reporting. All scalers and statistical baselines are fitted on the training window only. Because the validation and test windows lie entirely in the GP--HMM continuation, the reported test errors measure one-step-ahead accuracy on generator output, not on observed revenue.

\subsection{Lag construction}\label{sec:lags}
For forecasting, we transform the sequential problem into a supervised learning task via lagged inputs. For a univariate process $\{x_t\}$, each input is the lag vector $(x_{t-k},\dots,x_{t-1})$ and the target is the next value, $y_t = x_t$. Classical and quantum models therefore learn temporal dependence from the same lag structure. The lag length must balance expressiveness and parsimony: too few lags may omit relevant information, whereas too many produce high-dimensional, potentially collinear inputs and increase the risk of overfitting.

\begin{equation}
  \tilde{x}_t = \frac{x_t - x_{\min}}{x_{\max} - x_{\min}},
\end{equation}
The series is scaled with the scikit-learn \texttt{MinMaxScaler} \citep{sklearn}. Where $x_{\min}$ and $x_{\max}$ are the minimum and maximum of the training window (months 1--739). The scaler is fitted on the training window only and then applied unchanged to the validation and test periods, so values in these periods may fall slightly outside $[0,1]$.

The scaled series is converted into input--target pairs with input $(\tilde{x}_{t-4},\dots,\tilde{x}_{t-1})$ and target $\tilde{x}_t$, giving 735 pairs in the training window. Lags are constructed chronologically, so future values are never used as inputs. All models receive the same lag matrix: the recurrent models read it as a length-4 sequence of scalars, and the reservoirs as one 4-dimensional input vector per step.

\subsection{State preparation and amplitude encoding}
Classical inputs must be encoded as quantum states \citep{schuld2018,maheshwari2022}. We use amplitude encoding \citep{mottonen2005,pennylane,qalchemy}: for $\mathbf{x}\in\mathbb{R}^{2^n}$ the normalized vector $\tilde{\mathbf x}=\mathbf x/\|\mathbf x\|$ is prepared as
\begin{equation}
\ket{\psi_{\mathbf x}}=\sum_{i=0}^{2^n-1}\tilde x_i\ket{i},\qquad \sum_i|\tilde x_i|^2=1,
\end{equation}
so that a $2^n$-dimensional vector occupies $n$ qubits. We use exact statevector simulation for all circuits (Section~\ref{sec:setup}), so the amplitude-encoded state is prepared ideally.

\subsection{Classical LSTM}
The LSTM cell at time $t$ maintains a memory cell $c_t$ and hidden state $h_t$ with forget, input, candidate and output gates (Table~\ref{tab:eq}, left) \citep{hochreiter1997}, where $v_t$ is the concatenation of the input and the previous hidden state. We use a single layer with hidden size 256 and a linear output layer.

\begin{table}[h!]
\centering
\caption{LSTM and QLSTM gate equations.}
\label{tab:eq}
\renewcommand{\arraystretch}{1.1}
\begin{tabular}{>{\RaggedRight\arraybackslash}p{0.43\linewidth}|>{\RaggedRight\arraybackslash}p{0.43\linewidth}}
\toprule
\textbf{Classical LSTM} & \textbf{QLSTM} \\
\midrule
$f_t=\sigma(W_f v_t+b_f)$ & $f_t=\sigma(W_p\,\mathrm{VQC}_f(W_{\mathrm{in}}v_t))$ \\
$i_t=\sigma(W_i v_t+b_i)$ & $i_t=\sigma(W_p\,\mathrm{VQC}_i(W_{\mathrm{in}}v_t))$ \\
$\tilde C_t=\tanh(W_C v_t+b_C)$ & $\tilde C_t=\tanh(W_p\,\mathrm{VQC}_C(W_{\mathrm{in}}v_t))$ \\
$o_t=\sigma(W_o v_t+b_o)$ & $o_t=\sigma(W_p\,\mathrm{VQC}_o(W_{\mathrm{in}}v_t))$ \\
$c_t=f_t\odot c_{t-1}+i_t\odot\tilde C_t$ & $c_t=f_t\odot c_{t-1}+i_t\odot\tilde C_t$ \\
$h_t=o_t\odot\tanh(c_t)$ & $h_t=o_t\odot\tanh(c_t)$ \\
\bottomrule
\end{tabular}
\end{table}

\begin{figure}[h!]
\centering
\subfloat[LSTM]{\includegraphics[width=0.46\linewidth, trim=20 20 20 20, clip]{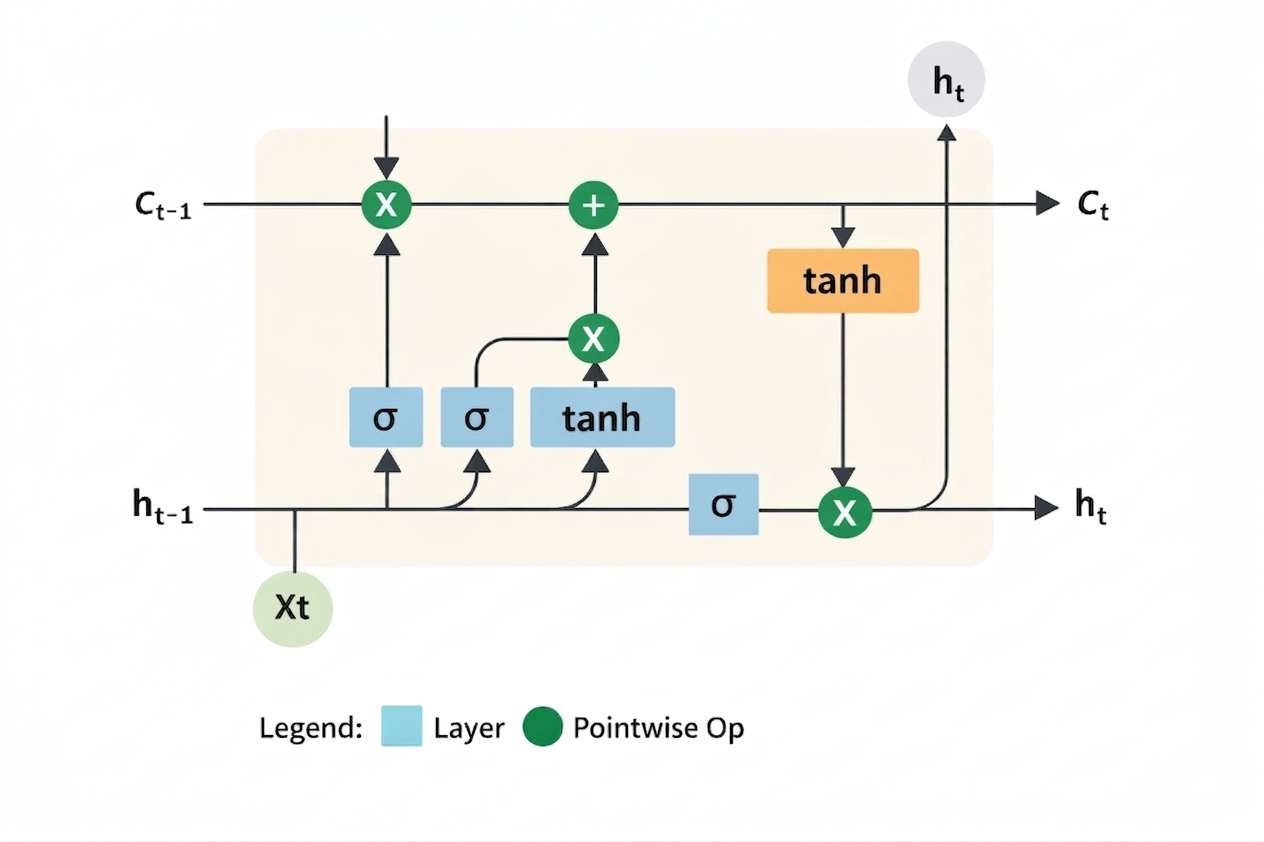}\label{fig:lstm}}
\hfill
\subfloat[QLSTM with VQC gates]{\includegraphics[width=0.46\linewidth, trim=10 10 10 10, clip]{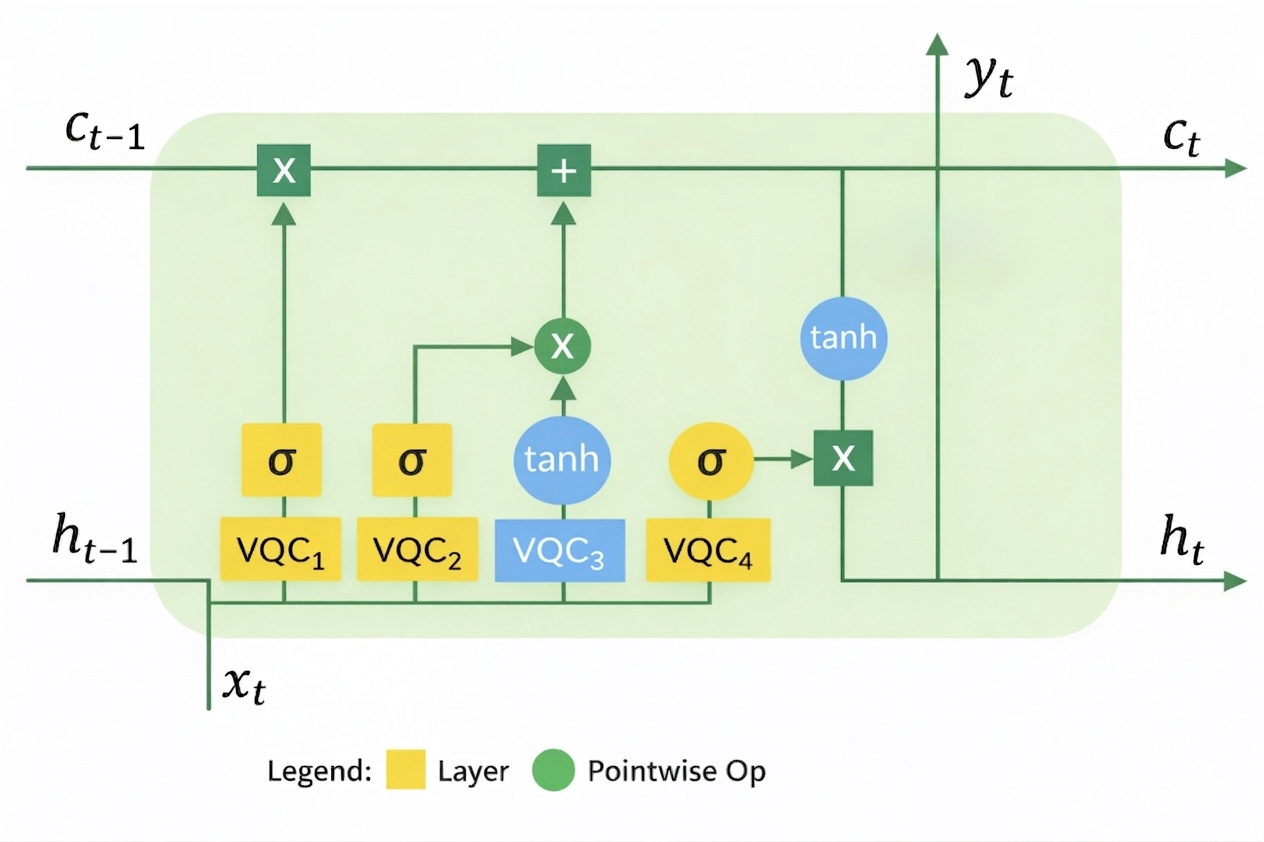}\label{fig:qlstm}}
\caption{Classical LSTM cell and the hybrid QLSTM cell in which each gate's affine map is replaced by a variational quantum circuit.}
\label{fig:lstm_combined}
\end{figure}

\subsection{Quantum LSTM}
\begin{figure}[h!]
\centering
\includegraphics[width=0.8\textwidth]{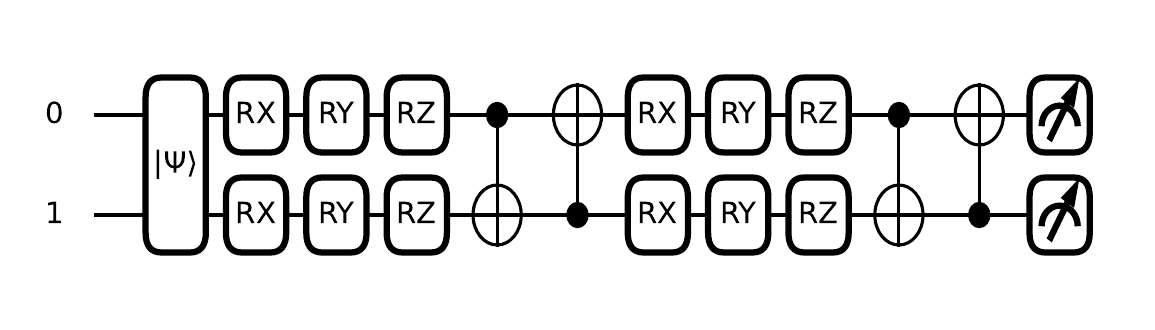}
\caption{The 2-qubit, 2-layer VQC used for every QLSTM gate: amplitude encoding, $R_X R_Y R_Z$ rotations per qubit and layer, ring CNOTs, Pauli-$Z$ readout (12 trainable angles).}
\label{fig:vqc}
\end{figure}

The QLSTM \citep{chen2022qlstm,maheshwari} replaces each gate's affine map with a VQC (Table~\ref{tab:eq}, right; Figure~\ref{fig:qlstm}). A classical interface layer $W_{\mathrm{in}}:\mathbb{R}^{1+256}\to\mathbb{R}^{4}$ maps $v_t$ to the amplitude-encoding dimension of $n=2$ qubits; each of the four gate VQCs consists of $L=2$ layers of $R_X,R_Y,R_Z$ rotations on every qubit followed by a ring of CNOTs (Figure~\ref{fig:vqc}), and returns the two Pauli-$Z$ expectations $\langle Z_0\rangle,\langle Z_1\rangle$; a second interface layer $W_p:\mathbb{R}^2\to\mathbb{R}^{256}$, shared by the four gates, maps the measurements to the hidden size. Each VQC has $L\times n\times 3=12$ trainable angles, i.e.\ $48$ quantum parameters in total; the interface and output layers hold the remaining 2,057 parameters (Table~\ref{tab:params}). The QLSTM is thus a hybrid model in which the quantum component is $2.3\%$ of the parameter budget, which is precisely why a parameter-matched classical control is required. Gradients through the circuits are exact (Section~\ref{sec:setup}).

\subsection{Parameter-matched classical control (cLSTM-S)}\label{sec:control}
To attribute any parameter-efficiency result to the quantum circuits rather than to the small interface architecture, we train \textbf{cLSTM-S}: a model identical to the QLSTM in which each 2-qubit VQC is replaced by a classical block with the same input and output dimensions, $\mathbf{u}\mapsto\tanh(A\,\mathbf u/\|\mathbf u\|+\mathbf b)$ with $A\in\mathbb{R}^{2\times4}$. The block keeps the same $L_2$ normalization the amplitude encoding applies and produces outputs in $[-1,1]$ like a Pauli-$Z$ expectation, so the only difference is the map between normalization and readout. Each block has 10 parameters, so cLSTM-S has 2,097 parameters versus 2,105 for the QLSTM; all interface layers, hidden size, optimizer, learning rate, early-stopping rule, augmentation and seeds are shared. If cLSTM-S matches the QLSTM within seed variance, the quantum components contribute no measurable benefit at this budget; if the QLSTM is better, the difference is attributable to the VQCs. As a second reference, we train \textbf{LSTM-S}, a plain LSTM with hidden size 21 (2,038 parameters) that has the small budget but not the QLSTM's interface structure.

\begin{table}[t]
\centering
\caption{Trainable-parameter counts.}
\label{tab:params}
\tabstyle
\begin{tabular}{lrrr}
\toprule
Model & Quantum & Classical & Total \\
\midrule
LSTM & 0 & 265,473 & 265,473 \\
QLSTM & 48 & 2,057 & 2,105 \\
cLSTM-S (matched control) & 0 & 2,097 & 2,097 \\
LSTM-S (small LSTM) & 0 & 2,038 & 2,038 \\
RC (linear readout) & 0 & 13 & 13 \\
RC (NN readout) & 0 & 225 & 225 \\
QRC (linear readout) & 0 & 13 & 13 \\
QRC (NN readout) & 0 & 225 & 225 \\
\bottomrule
\end{tabular}

\end{table}

\subsection{Reservoir computing}
\begin{figure}[h!]
\centering
\subfloat[RC]{\includegraphics[width=0.40\linewidth, trim=20 20 20 20, clip]{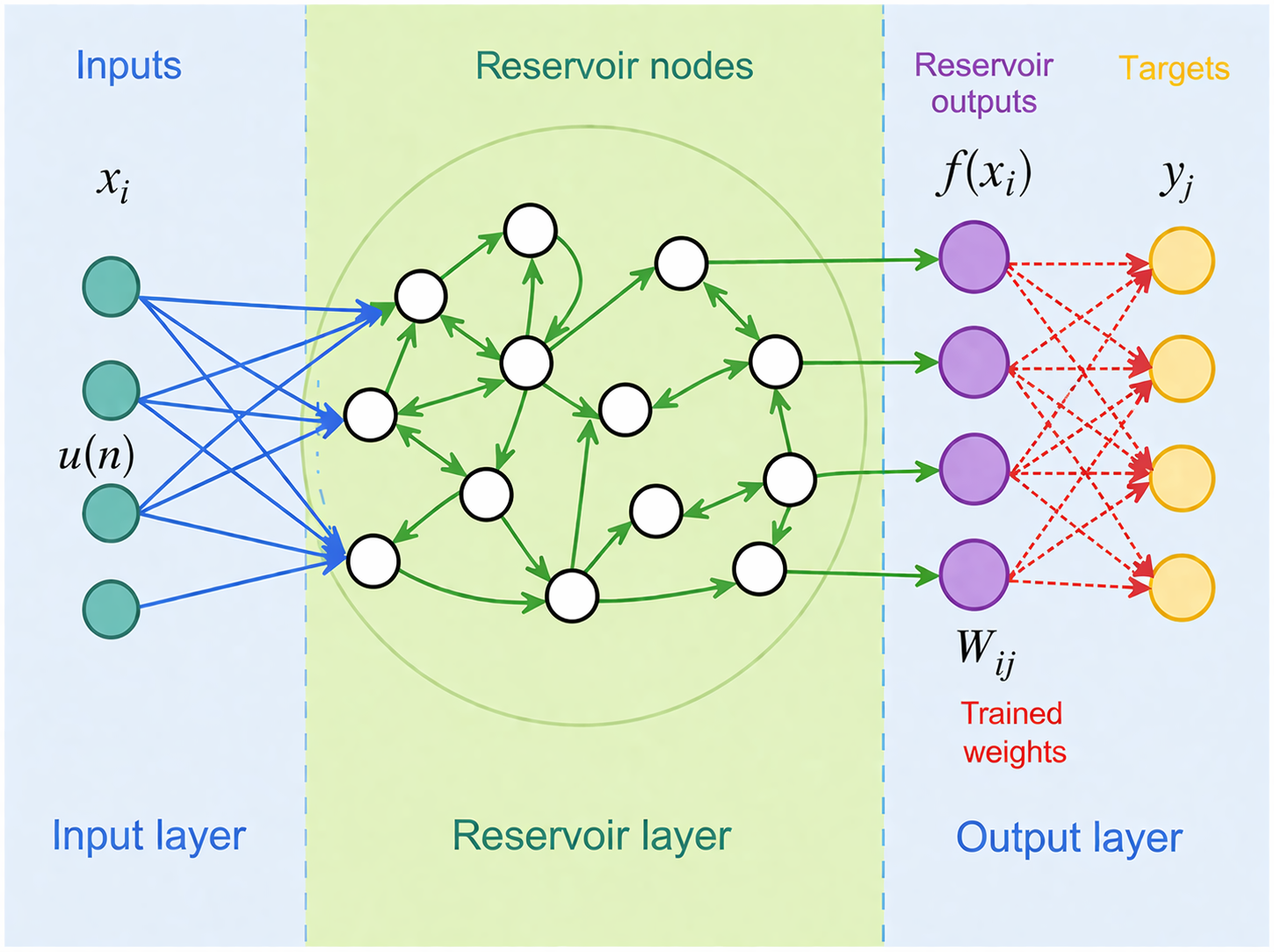}\label{fig:rc}}
\hfill
\subfloat[QRC]{\includegraphics[width=0.40\linewidth, trim=10 10 10 10, clip]{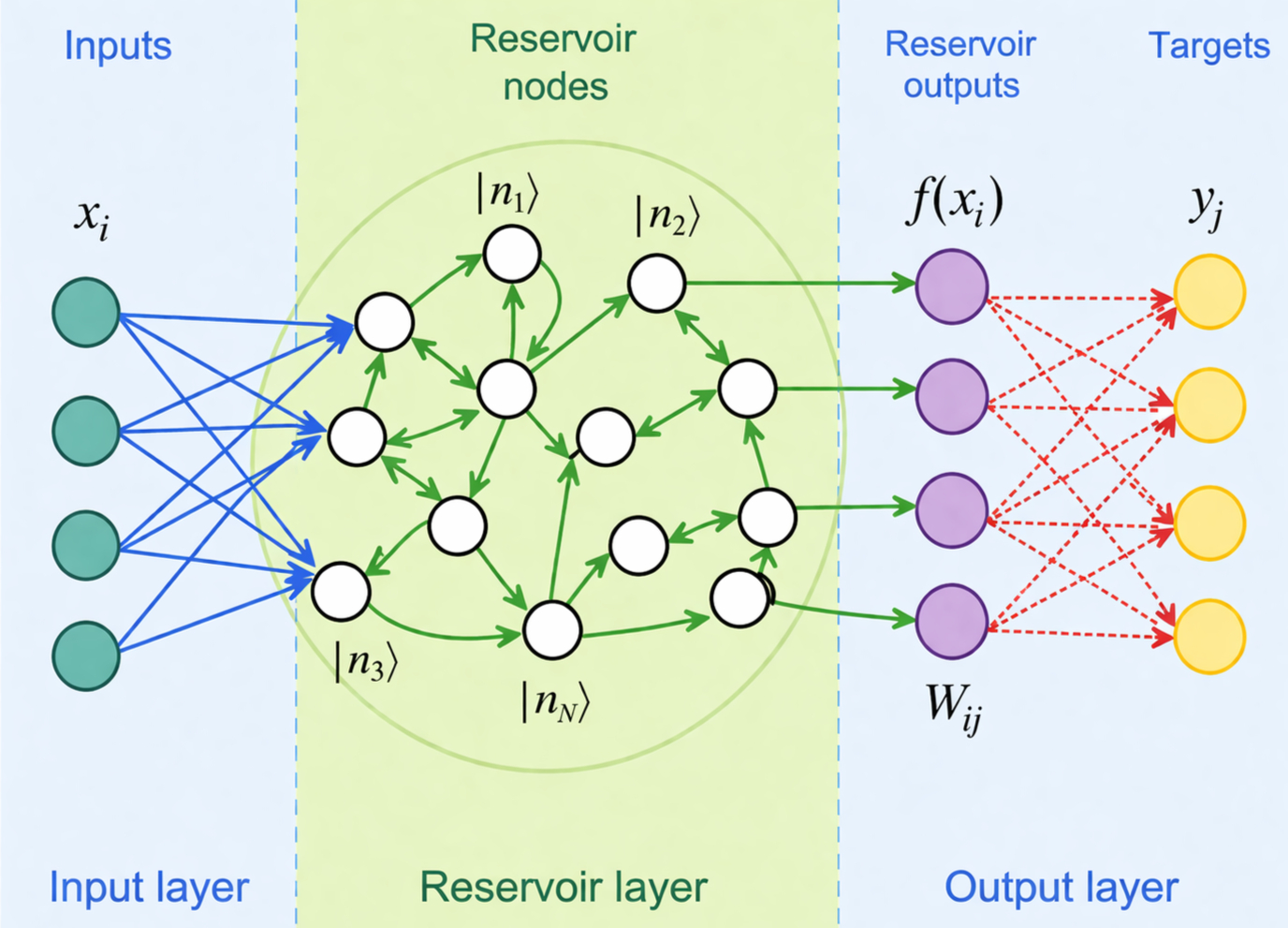}\label{fig:qrc}}
\caption{Classical reservoir computing and quantum reservoir computing; in both cases only the readout is trained.}
\label{fig:rc_combined}
\end{figure}

In RC, only a linear readout is trained; the recurrent reservoir is fixed after random initialization \citep{jaeger2001} (Figure~\ref{fig:rc}). We use a leaky echo-state network with $N=12$ tanh units, $r_t=(1-\lambda)r_{t-1}+\lambda\tanh(W_{\mathrm{in}}x_t+Wr_{t-1}+b)$, spectral radius $0.9$, $\lambda=0.5$, input weights and biases uniform in $[-0.5,0.5]$. $N=12$ equals the feature dimension $3n$ of the quantum reservoir below.

\subsection{Quantum reservoir computing}
QRC uses a fixed, randomly initialized quantum circuit as the reservoir (Figure~\ref{fig:qrc}) \citep{fujii2017,ghosh2019,maheshwari}. Rather than zero-padding inputs to length $2^n$, which biases the state and wastes Hilbert-space capacity, we apply a fixed random projection followed by a single normalization. Given the lag vector $x^{(t)}\in\mathbb{R}^{d_x}$ ($d_x=4$), a bias $b\in\mathbb{R}^n$, and the previous measurement vector $r^{(t-1)}\in\mathbb{R}^{3n}$ \citep{jaeger2001}:
\begin{align}
x^{(t)}_i&=x^{(t)}_{i\bmod d_x}+b_i,\quad i=1,\dots,n,\\
m^{(t-1)}_i&=w_X\langle X_i\rangle^{(t-1)}+w_Y\langle Y_i\rangle^{(t-1)}+w_Z\langle Z_i\rangle^{(t-1)},\\
v^{(t)}_i&=\lambda x^{(t)}_i+(1-\lambda)m^{(t-1)}_i,\qquad \lambda\in[0,1],\\
z^{(t)}&=Pv^{(t)},\quad P\in\mathbb{R}^{2^n\times n},\quad \tilde z^{(t)}=\frac{z^{(t)}}{\|z^{(t)}\|_2},\\
\ket{\psi^{(t)}_0}&=\textstyle\sum_{j=0}^{2^n-1}\tilde z^{(t)}_j\ket{j}.
\end{align}
For each layer $l=1,\dots,L$ we apply random single-qubit rotations and ring entanglement,
\begin{equation}
U^{(l)}=\Big(\prod_{i=1}^{n}R(\theta_{l,i},\phi_{l,i},\lambda_{l,i})_i\Big)\Big(\prod_{i=1}^{n-1}\mathrm{CNOT}_{i\to i+1}\Big)\mathrm{CNOT}_{n\to1},
\end{equation}
and $\ket{\psi^{(t)}}=U^{(L)}\cdots U^{(1)}\ket{\psi^{(t)}_0}$. The feature vector stacks $\langle X_i\rangle,\langle Y_i\rangle,\langle Z_i\rangle$ for all qubits, $r^{(t)}\in\mathbb{R}^{3n}$, and the linear prediction is $\hat y^{(t)}=W_{\mathrm{out}}^\top r^{(t)}$ with ridge solution
\begin{equation}
W_{\mathrm{out}}=(R^\top R+\lambda_{\mathrm{reg}}I)^{-1}R^\top Y.
\end{equation}
We use $n=4$, $L=3$, $\lambda=0.5$, $w_X=w_Y=w_Z=1/3$, $\lambda_{\mathrm{reg}}=10^{-6}$, $P_{ij}\sim\mathcal N(0,1/n)$, $b_i\sim\mathcal U(-0.5,0.5)$. The classical RC is matched to the same feature dimension ($3n=12$) and the same readouts. Both reservoirs are evaluated with the ridge readout (13 parameters including bias) and with a neural-network (NN) readout, a one-hidden-layer ReLU network with 16 units (225 parameters) trained with Adam (learning rate $10^{-2}$) and early stopping on the validation slice. All reservoir hyperparameters ($n$, $L$, $\lambda$, spectral radius, $\lambda_{\mathrm{reg}}$) were fixed a priori and not tuned on validation data; this is a limitation noted in Section~\ref{sec:results}. The series is run once, sequentially, through the reservoir over months 1--1056 (and the input for month 1057); the readouts are fitted on the reservoir features of the training window.

\subsection{Benchmark forecasts}\label{sec:benchmarks}
For monthly revenue data with possible seasonality, we include three standard benchmark forecasts, also called baseline models \citep{hyndman2021}, fitted or evaluated on the training window only and scored on the same 159 test months:
\begin{itemize}
\item \textbf{Persistence}: $\hat x_t=x_{t-1}$.
\item \textbf{Seasonal-naive}: $\hat x_t=x_{t-12}$.
\item \textbf{SARIMA}$(p,d,q)(P,D,Q)_{12}$: orders selected by AIC on the training window over $p,q\in\{0,1,2\}$, $d\in\{0,1\}$, $(P,D,Q)\in\{(0,0,0),(1,0,0),(0,1,0),(0,1,1),(1,0,1)\}$, fitted with \texttt{statsmodels} \citep{statsmodels}; one-step-ahead test forecasts use the training-window parameters held fixed.
\end{itemize}
We regard a learned model as skilful only if its test RMSE is below that of the best baseline model, consistently over replications.

\subsection{Experimental setup and statistical analysis}\label{sec:setup}
All models are implemented in Python~3 with PyTorch, NumPy, scikit-learn \citep{sklearn} and \texttt{statsmodels} \citep{statsmodels}. The 2-qubit gate circuits of the QLSTM are simulated exactly as statevectors in PyTorch, so that automatic differentiation yields exact gradients; the implementation was verified to agree with PennyLane's \texttt{default.qubit} simulator \citep{pennylane} to $2\times10^{-7}$ on random inputs and parameters. The 4-qubit QRC unitary is simulated exactly in NumPy. All runs use a single CPU core, and no hardware noise is modelled.

Every model is run with 10 seeds ($0,\dots,9$), which control weight initialization, mini-batch order and reservoir construction, giving 10 independent runs per model. All models make one-step-ahead forecasts: the input for test month $t$ consists of the four preceding months of the series. LSTM, QLSTM, cLSTM-S and LSTM-S share the Adam optimizer, learning rate $10^{-3}$, MSE loss, shuffled mini-batches of 32 samples, at most 300 epochs, early stopping on the validation window with patience 20, and restoration of the best-validation checkpoint. Reservoir readouts are fitted as described above.

For each model we report the mean and standard deviation of the test error over the 10 replications; the baseline forecasts are deterministic. Besides MSE, RMSE and MAE we report the mean absolute scaled error (MASE), i.e.\ the test MAE divided by the in-sample MAE of the seasonal-naive forecast on the training window, and the number of seeds for which a model has a lower test RMSE than SARIMA. With a single series, the unit of replication is the seed. Paired comparisons use the Wilcoxon signed-rank test over the 10 seeds (models with the same seed are paired; the smallest attainable two-sided $p$-value is $0.002$); the three primary comparisons (LSTM vs.\ QLSTM, QLSTM vs.\ cLSTM-S, QRC vs.\ RC with linear readout), fixed before the experiments were run, are Holm-corrected. Comparisons with SARIMA use the one-sample Wilcoxon test of the per-seed RMSE difference.

\section{Results and Discussion}\label{sec:results}

In this study, we use both classical and quantum algorithms for time-series prediction. The training models
are implemented in Python3 using PyTorch, scikit-learn, and an H200 NVIDIA GPU. We access Q-Alchemy via its API
to encode classical data into quantum states and integrate this encoding into the PennyLane framework.
We compare the performance and capabilities of QLSTM and QRC with their classical counterparts across
univariate time-series prediction tasks.

\begin{figure}[h!]
\centering
\includegraphics[width=\textwidth]{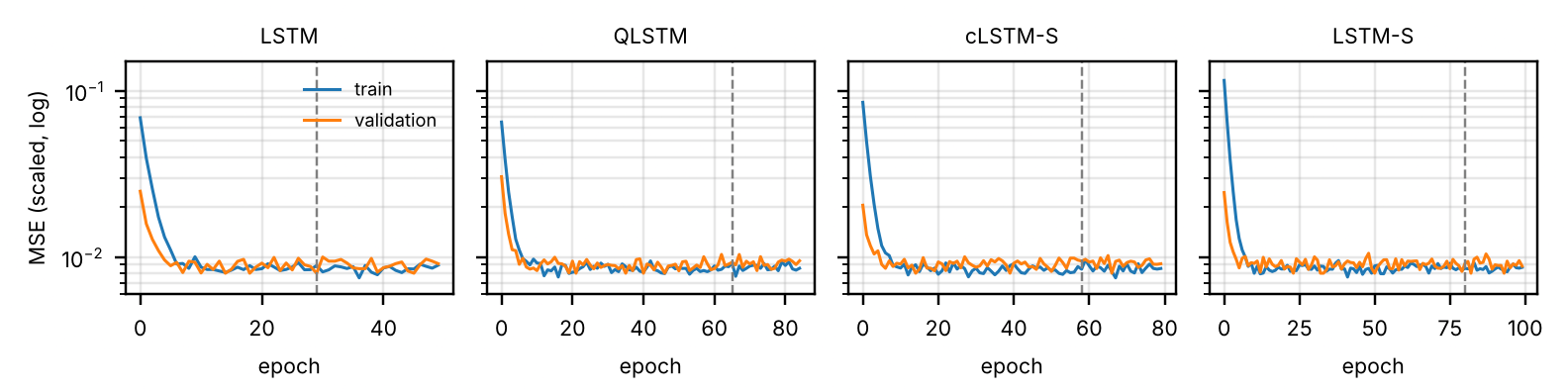}
\caption{Training and validation loss (MSE) for LSTM, QLSTM, cLSTM-S and LSTM-S, each at the seed with median QLSTM test RMSE. The dashed line marks the epoch of the restored best-validation checkpoint, not the epoch at which training stopped.}
\label{fig:loss}
\end{figure}

\begin{figure}[h!]
\centering
\includegraphics[width=\textwidth]{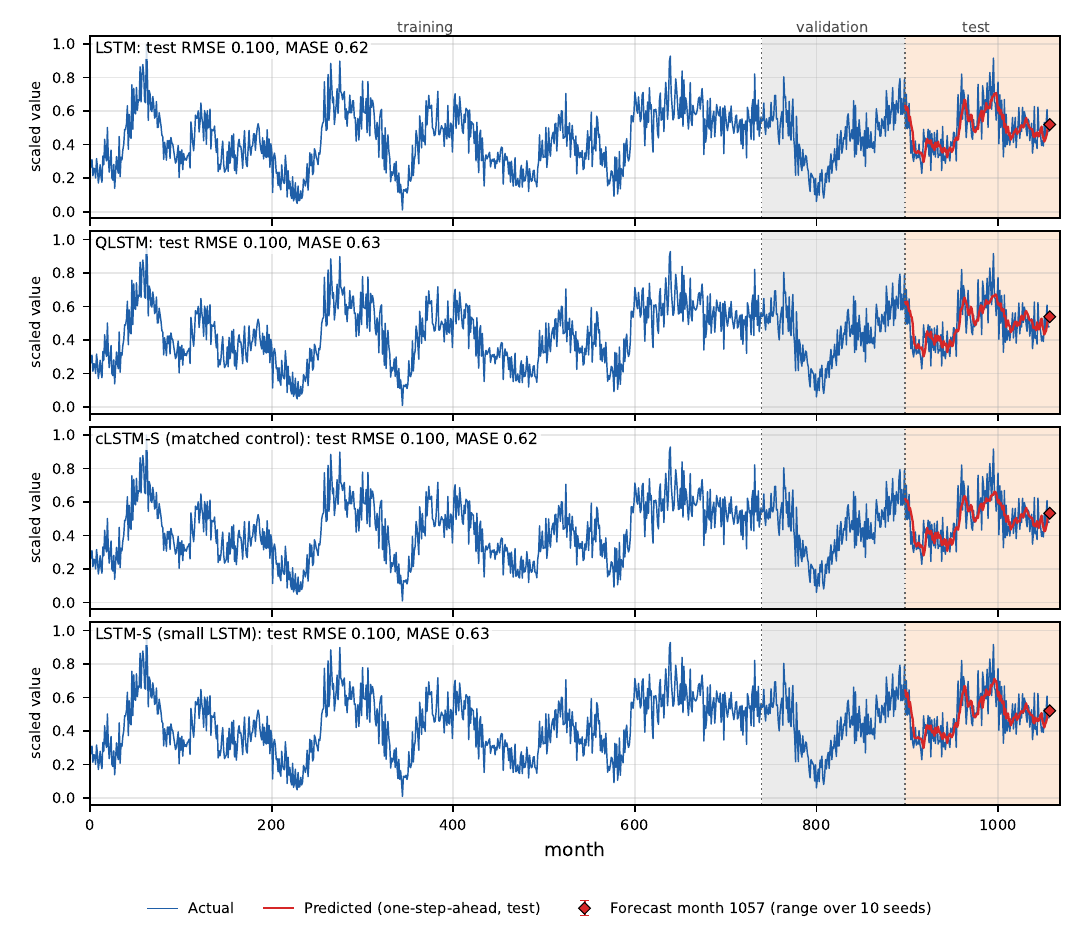}
\caption{Recurrent models (median-RMSE QLSTM seed), one panel per model: actual values (blue) in the training, validation and test windows, one-step-ahead test predictions (red), and the one-step-ahead forecast of month 1057 (mean and range over 10 seeds; this month has no observation and is not scored). Panel titles give the test RMSE and MASE of the plotted seed.}
\label{fig:traj_rec}
\end{figure}

\begin{table}[h!]
\centering
\caption{Recurrent-model test error on the 159 test months (mean $\pm$ std over 10 seeds; models trained on months 1--739). Last column: seeds with lower test RMSE than SARIMA.}
\label{tab:rec}
\footnotesize
\setlength{\tabcolsep}{3pt}
\resizebox{\linewidth}{!}{\begin{tabular}{lrccccc}
\toprule
Model & Params & MSE & RMSE & MAE & MASE & Seeds $<$ SARIMA \\
\midrule
LSTM & 265,473 & 0.00996 $\pm$ 0.00002 & 0.0998 $\pm$ 0.0001 & 0.0805 $\pm$ 0.0003 & 0.625 $\pm$ 0.002 & 10/10 \\
QLSTM & 2,105 & 0.01001 $\pm$ 0.00015 & 0.1001 $\pm$ 0.0007 & 0.0804 $\pm$ 0.0007 & 0.624 $\pm$ 0.005 & 9/10 \\
cLSTM-S (matched control) & 2,097 & 0.01000 $\pm$ 0.00006 & 0.1000 $\pm$ 0.0003 & 0.0802 $\pm$ 0.0005 & 0.623 $\pm$ 0.004 & 10/10 \\
LSTM-S (small LSTM) & 2,038 & 0.01001 $\pm$ 0.00003 & 0.1000 $\pm$ 0.0001 & 0.0807 $\pm$ 0.0001 & 0.627 $\pm$ 0.001 & 10/10 \\
\bottomrule
\end{tabular}
}
\end{table}
Table~\ref{tab:rec} summarizes the recurrent-model comparison, and Table~\ref{tab:stats} reports paired tests. All four recurrent models reach essentially the same test error: RMSE $0.0998\pm0.0001$ for the full-size LSTM, $0.1001\pm0.0007$ for the QLSTM, $0.1000\pm0.0003$ for cLSTM-S and $0.1000\pm0.0001$ for LSTM-S (MASE $0.623$--$0.627$). None of the paired differences is significant: LSTM versus QLSTM $-0.0003$ ($p=0.28$, Holm-adjusted $p=0.55$), QLSTM versus cLSTM-S $+0.00005$ ($p=0.92$), QLSTM versus LSTM-S $+0.00003$ ($p=0.49$); the MASE-based tests agree. Each recurrent model is better than SARIMA in at least 9 of 10 seeds ($p\le0.004$, Table~\ref{tab:vsnull}), by about $0.002$ RMSE or 2\%. The QLSTM shows the largest seed-to-seed variability (std $0.0007$). Figure~\ref{fig:traj_rec} shows that all four models follow the movements of the series in the test window; the standard deviation of their one-step-ahead predictions is $0.093$--$0.096$, compared with $0.132$ for the actual test values. Their one-step-ahead forecasts of month 1057 lie between $0.52$ and $0.55$ (scaled), close to the SARIMA forecast of $0.50$. The loss curves in Figure~\ref{fig:loss} show early stopping after 63--137 epochs on average (Table~\ref{tab:cost}).

These results show that the QLSTM matches the full-size LSTM with 126 times fewer parameters. However, cLSTM-S and LSTM-S, both with about 2,100 parameters and no quantum circuit, match it as well. The parameter efficiency of the QLSTM is therefore shared by classical models of the same size and cannot be attributed to the quantum circuits. This conclusion applies to the tested VQC ansatz and to this normalized linear-$\tanh$ control; alternative classical blocks or circuit ans\"atze could yield different results.

\begin{table}[h!]
\centering
\caption{Reservoir-model test error (same conventions as Table~\ref{tab:rec}); seeds control the reservoir construction.}
\label{tab:res}
\footnotesize
\setlength{\tabcolsep}{3pt}
\resizebox{\linewidth}{!}{\begin{tabular}{lrccccc}
\toprule
Model & Params & MSE & RMSE & MAE & MASE & Seeds $<$ SARIMA \\
\midrule
RC (linear readout) & 13 & 0.01002 $\pm$ 0.00010 & 0.1001 $\pm$ 0.0005 & 0.0797 $\pm$ 0.0004 & 0.619 $\pm$ 0.003 & 10/10 \\
RC (NN readout) & 225 & 0.00996 $\pm$ 0.00018 & 0.0998 $\pm$ 0.0009 & 0.0799 $\pm$ 0.0008 & 0.621 $\pm$ 0.006 & 10/10 \\
QRC (linear readout) & 13 & 0.01395 $\pm$ 0.00287 & 0.1176 $\pm$ 0.0116 & 0.0935 $\pm$ 0.0079 & 0.726 $\pm$ 0.061 & 0/10 \\
QRC (NN readout) & 225 & 0.01381 $\pm$ 0.00272 & 0.1170 $\pm$ 0.0110 & 0.0928 $\pm$ 0.0079 & 0.721 $\pm$ 0.061 & 0/10 \\
\bottomrule
\end{tabular}
}
\end{table}
Table~\ref{tab:res} compares the four reservoir variants. The classical reservoir reaches the level of the recurrent models: RMSE $0.1001\pm0.0005$ with the ridge readout and $0.0998\pm0.0009$ with the NN readout, better than SARIMA in all 10 seeds ($p=0.002$). The quantum reservoir is clearly worse: RMSE $0.1176\pm0.0116$ (ridge) and $0.1170\pm0.0110$ (NN), worse than SARIMA in all 10 seeds ($p=0.002$). RC is better than QRC in all 10 seeds with both readouts ($p=0.002$; Holm-adjusted $p=0.006$ for the linear readout), by $0.017$ RMSE. The QRC error also varies about ten times more across reservoir seeds than the RC error, i.e.\ the quality of the random quantum reservoir depends strongly on its construction. The choice of readout does not matter for either reservoir ($p\ge0.16$).

Because the feature dimension and the readouts are matched, this gap reflects the reservoir features themselves: the echo-state features gave lower errors than the quantum-reservoir measurement features in every seed. Reservoir size, circuit depth, input mixing, and readout regularization were fixed a priori. Validation-tuned reservoir hyperparameters, or selecting the reservoir on the validation window, are natural directions for improving the QRC and could change the size of the gap.

\begin{figure}[h!]
\centering
\includegraphics[width=\textwidth]{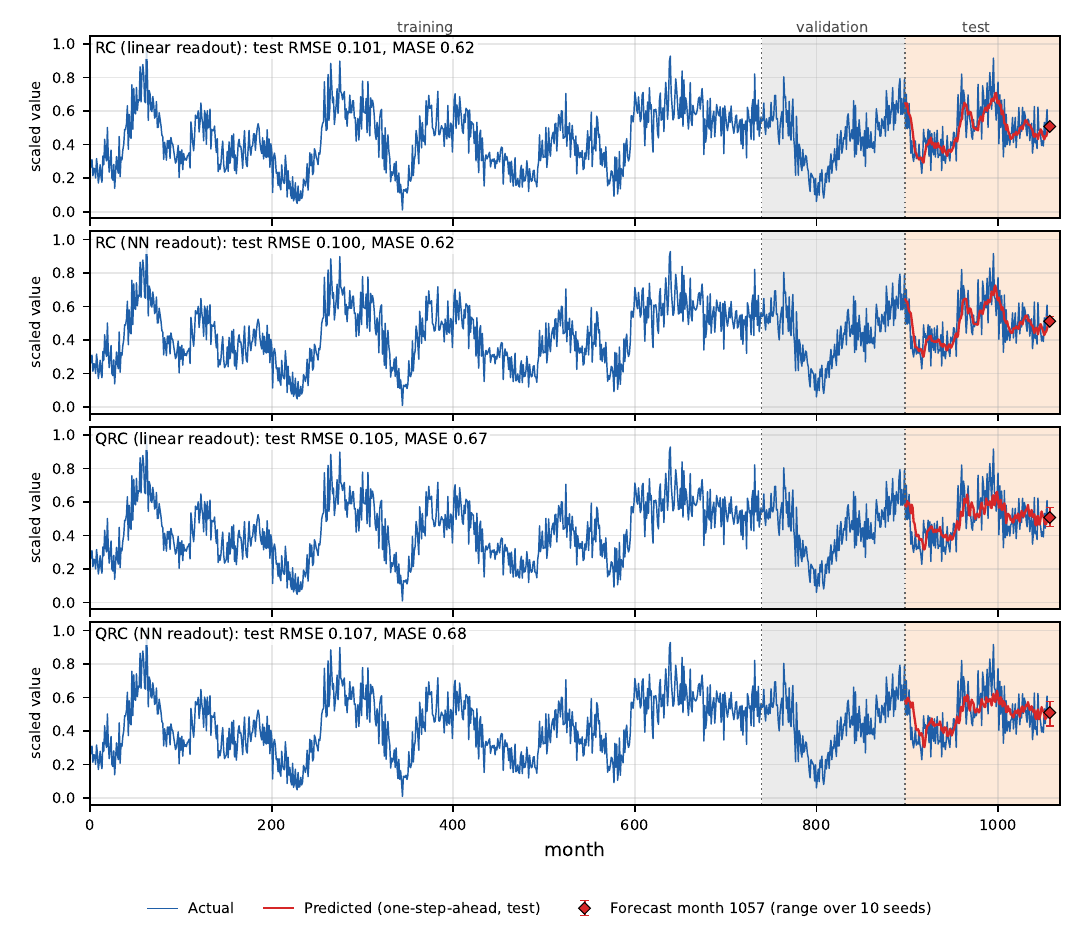}
\caption{As Figure~\ref{fig:traj_rec} for the four reservoir variants: training, validation and test actuals, one-step-ahead test predictions and the forecast of month 1057.}
\label{fig:traj_res}
\end{figure}

\begin{table}[h!]
\centering
\caption{Baseline test error on the 159 test months (months 898--1056, scaled units); the baselines are deterministic.}
\label{tab:null}
\tabstyle
\begin{tabular}{lcccc}
\toprule
Model & MSE & RMSE & MAE & MASE \\
\midrule
Persistence & 0.01830 & 0.1353 & 0.1099 & 0.854 \\
Seasonal-naive & 0.02784 & 0.1669 & 0.1331 & 1.034 \\
SARIMA & 0.01036 & 0.1018 & 0.0811 & 0.630 \\
\bottomrule
\end{tabular}

\end{table}
Table~\ref{tab:null} reports the baseline forecasts. The AIC-selected SARIMA, of order $(2,0,2)$ without a seasonal term, is the strongest baseline (RMSE $0.1018$, MASE $0.630$); persistence (RMSE $0.1353$) and seasonal-naive (RMSE $0.1669$) are clearly weaker. The SARIMA RMSE of $0.1018$ is the reference against which every learned model is measured.

\begin{figure}[h!]
\centering
\includegraphics[width=\textwidth]{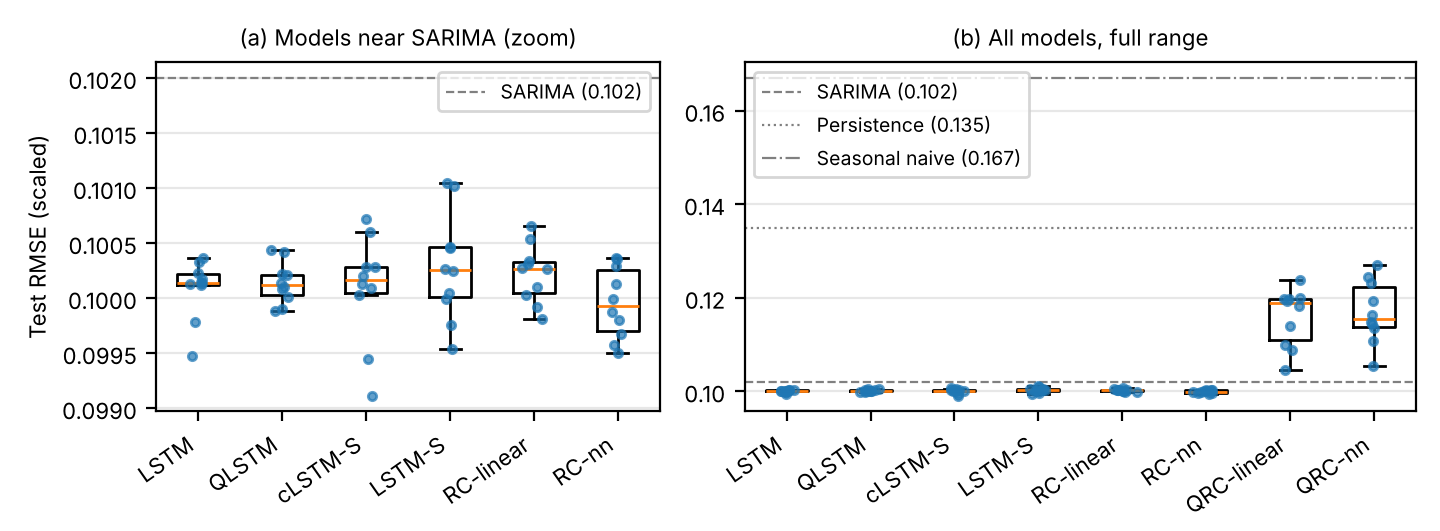}
\caption{Test RMSE (scaled) of the learned models over 10 random seeds. Boxes show the median and interquartile range, points show individual seeds, and triangles mark the means. Horizontal lines show the baseline forecasts: SARIMA (0.102), persistence (0.135) and seasonal naive (0.167). \textbf{(A)} Models within 0.003 of SARIMA, on a zoomed y-axis. \textbf{(B)} All models on the full y-axis range.}
\label{fig:rel}
\end{figure}

\begin{table}[t]
\centering
\caption{Paired Wilcoxon signed-rank tests on test RMSE (A minus B; negative favours A). ``A wins'': seeds for which A has the lower test RMSE; Wilcoxon tests over the 10 seeds. Holm-corrected $p$-values for the three primary comparisons in brackets.}
\label{tab:stats}
\tabstyle
\begin{tabular}{llrccc}
\toprule
A & B & $\overline{\Delta\mathrm{RMSE}}$ & A wins & $p$ (RMSE) & $p$ (MASE) \\
\midrule
LSTM & QLSTM & -0.0003 & 7/10 & 0.275 [0.551] & 0.275 \\
QLSTM & cLSTM-S & +0.0000 & 6/10 & 0.922 [0.922] & 0.625 \\
QRC-linear & RC-linear & +0.0175 & 0/10 & 0.002 [0.006] & 0.002 \\
\midrule
QLSTM & LSTM-S & +0.0000 & 5/10 & 0.492 & 0.131 \\
LSTM & cLSTM-S & -0.0002 & 7/10 & 0.084 & 0.064 \\
QRC-nn & RC-nn & +0.0172 & 0/10 & 0.002 & 0.002 \\
QRC-linear & QRC-nn & +0.0006 & 2/10 & 0.160 & 0.160 \\
RC-linear & RC-nn & +0.0003 & 4/10 & 0.375 & 0.557 \\
\bottomrule
\end{tabular}

\end{table}
\begin{table}[t]
\centering
\caption{Learned models versus SARIMA, the best benchmark (one-sample Wilcoxon test of the per-seed RMSE difference, 10 seeds).}
\label{tab:vsnull}
\tabstyle
\begin{tabular}{lrcc}
\toprule
Model & $\overline{\Delta\mathrm{RMSE}}$ vs.\ SARIMA & Seeds better & $p$ (Wilcoxon, $n=10$) \\
\midrule
LSTM & -0.0020 & 10/10 & 0.002 \\
QLSTM & -0.0018 & 9/10 & 0.004 \\
cLSTM-S (matched control) & -0.0018 & 10/10 & 0.002 \\
LSTM-S (small LSTM) & -0.0018 & 10/10 & 0.002 \\
RC (linear readout) & -0.0017 & 10/10 & 0.002 \\
RC (NN readout) & -0.0020 & 10/10 & 0.002 \\
QRC (linear readout) & +0.0158 & 0/10 & 0.002 \\
QRC (NN readout) & +0.0152 & 0/10 & 0.002 \\
\bottomrule
\end{tabular}

\end{table}
Tables~\ref{tab:stats} and~\ref{tab:vsnull} summarize the paired tests, and Fig.~\ref{fig:rel} shows the distribution of the test RMSE over seeds together with the benchmark errors. Two conclusions hold for both RMSE and MASE: the QLSTM is statistically indistinguishable from the full-size LSTM and from both classical models of equal size, and RC is better than QRC.

All four recurrent models and both classical reservoirs are therefore skilful in the sense defined in Section~\ref{sec:benchmarks}, whereas the quantum reservoir is not. The margin over SARIMA, however, is only about 2\% in RMSE. On this series most of the predictable structure is captured by a low-order linear model, and the learned models add little to it. This agrees with the general finding of \citet{makridakis2018} for business series and with \citet{bowles2024} for small quantum models: the comparison to a well-chosen baseline is more informative than the comparison between learned models.

\begin{table}[t]
\centering
\caption{Training-window ablation: test RMSE (mean $\pm$ seed std) when models are trained on the first 96 months only (92 lag samples) versus the full training window (months 1--739, 735 lag samples); $\Delta$ = full minus first-96; $p$ from a paired Wilcoxon test over seeds.}
\label{tab:abl}
\tabstyle
\begin{tabular}{lccrc}
\toprule
Model & First 96 months & Months 1--739 & $\Delta$ & $p$ \\
\midrule
LSTM & 0.1005 $\pm$ 0.0005 & 0.0998 $\pm$ 0.0001 & -0.0007 & 0.002 \\
QLSTM & 0.1157 $\pm$ 0.0148 & 0.1001 $\pm$ 0.0007 & -0.0157 & 0.002 \\
cLSTM-S (matched control) & 0.1149 $\pm$ 0.0147 & 0.1000 $\pm$ 0.0003 & -0.0149 & 0.002 \\
LSTM-S (small LSTM) & 0.1006 $\pm$ 0.0001 & 0.1000 $\pm$ 0.0001 & -0.0005 & 0.002 \\
RC (linear readout) & 0.1025 $\pm$ 0.0015 & 0.1001 $\pm$ 0.0005 & -0.0025 & 0.004 \\
RC (NN readout) & 0.1005 $\pm$ 0.0017 & 0.0998 $\pm$ 0.0009 & -0.0007 & 0.027 \\
QRC (linear readout) & 0.1237 $\pm$ 0.0099 & 0.1176 $\pm$ 0.0116 & -0.0061 & 0.002 \\
QRC (NN readout) & 0.1189 $\pm$ 0.0098 & 0.1170 $\pm$ 0.0110 & -0.0018 & 0.084 \\
\bottomrule
\end{tabular}

\end{table}
Table~\ref{tab:abl} repeats the experiment with models trained on the first 96 months only (validation and test windows unchanged). The longer training window lowers the error of every model. The effect is small for the LSTM, LSTM-S and RC ($\le0.0025$ RMSE) but large for the two models with the narrow 4-dimensional gate bottleneck: the QLSTM improves from $0.1157\pm0.0148$ to $0.1001$ and cLSTM-S from $0.1149\pm0.0147$ to $0.1000$ ($p=0.002$ for both), and the variability across seeds drops by a factor of about 20. With 92 training samples the bottleneck models fail in some seeds; with 735 samples they reach the level of the other models.

To assess longer horizons, every trained model also produces 24-month forecasts by recursion: the one-step-ahead prediction is appended to the lag window and fed back as input, and the reservoirs are driven by their own predictions. No model is retrained for this purpose; each retrained model was verified to reproduce its logged one-step-ahead test predictions exactly. Because the months after 1056 are unobserved, accuracy is measured in a rolling-origin backtest inside the test window: from each origin $o=897,\dots,1032$ (136 origins) a 24-month forecast of months $o+1,\dots,o+24$ is made using only data up to month $o$, and the RMSE is computed for each horizon $h$ over all origins. SARIMA uses its native multi-step forecast with the training-window parameters, persistence repeats the last value, and seasonal-naive repeats the value 12 months earlier.

\begin{table}[h!]
\centering
\caption{Rolling-origin 24-month backtest (136 origins, months 898--1056): RMSE at horizons $h=1,6,12,24$ and over all horizons (mean $\pm$ std over 10 seeds). Last two columns: seeds with lower all-horizon RMSE than SARIMA and the one-sample Wilcoxon $p$-value.}
\label{tab:f24}
\footnotesize
\resizebox{\linewidth}{!}{\begin{tabular}{lccccccc}
\toprule
Model & $h=1$ & $h=6$ & $h=12$ & $h=24$ & $h=1$--$24$ & Seeds $<$ SARIMA & $p$ \\
\midrule
Persistence & 0.139 & 0.159 & 0.163 & 0.176 & 0.1652 & -- & -- \\
Seasonal-naive & 0.173 & 0.172 & 0.163 & 0.176 & 0.1780 & -- & -- \\
SARIMA & 0.104 & 0.134 & 0.166 & 0.176 & 0.1576 & -- & -- \\
\midrule
LSTM & 0.102 & 0.122 & 0.135 & 0.130 & 0.1273 $\pm$ 0.0014 & 10/10 & 0.002 \\
QLSTM & 0.102 & 0.123 & 0.139 & 0.139 & 0.1315 $\pm$ 0.0027 & 10/10 & 0.002 \\
cLSTM-S (matched control) & 0.102 & 0.123 & 0.140 & 0.142 & 0.1331 $\pm$ 0.0040 & 10/10 & 0.002 \\
LSTM-S (small LSTM) & 0.102 & 0.122 & 0.136 & 0.130 & 0.1274 $\pm$ 0.0007 & 10/10 & 0.002 \\
RC (linear readout) & 0.102 & 0.123 & 0.140 & 0.134 & 0.1313 $\pm$ 0.0031 & 10/10 & 0.002 \\
RC (NN readout) & 0.102 & 0.123 & 0.137 & 0.131 & 0.1285 $\pm$ 0.0025 & 10/10 & 0.002 \\
QRC (linear readout) & 0.121 & 0.137 & 0.145 & 0.151 & 0.1500 $\pm$ 0.0347 & 7/10 & 0.160 \\
QRC (NN readout) & 0.120 & 0.135 & 0.140 & 0.136 & 0.1374 $\pm$ 0.0065 & 10/10 & 0.002 \\
\bottomrule
\end{tabular}
}
\end{table}

\begin{figure}[h!]
\centering
\includegraphics[width=0.9\textwidth]{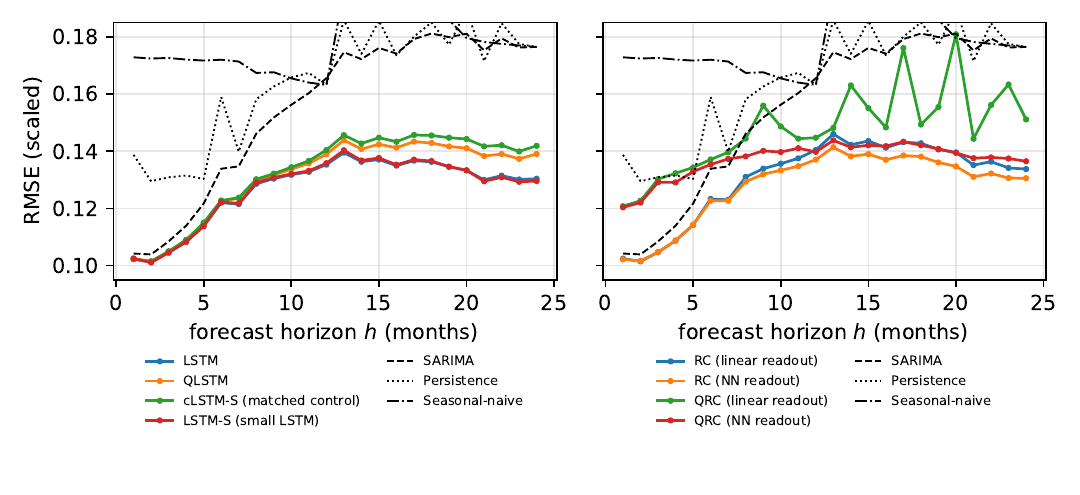}
\caption{Rolling-origin backtest: RMSE as a function of the forecast horizon (mean over 10 seeds; 136 origins in the test window) for the recurrent models \textbf{(A)} and the reservoirs \textbf{(B)}, with the three baseline forecasts.}
\label{fig:f24h}
\end{figure}

Table~\ref{tab:f24} and Figure~\ref{fig:f24h} show the 24-month backtest. One month ahead, the recurrent models and RC reach an RMSE of about $0.10$, SARIMA $0.104$, and QRC $0.12$. As the forecast gets longer, the error grows for every model. SARIMA gets worse fastest ($0.176$ at 24 months), while the learned models level off at about $0.13$--$0.14$.

Averaged over all horizons, the learned models beat SARIMA ($0.1576$) in all 10 seeds. The one exception is QRC with linear readout, which does so in only 7 of 10 seeds. The full-size LSTM ($0.1273$) and LSTM-S ($0.1274$) are the best. They beat QLSTM ($0.1315$) in all 10 seeds, whereas QLSTM and cLSTM-S perform alike ($p=0.38$). QRC is worse than RC, and with linear readout it is also unstable.

\begin{figure}[h!]
\centering
\includegraphics[width=\textwidth]{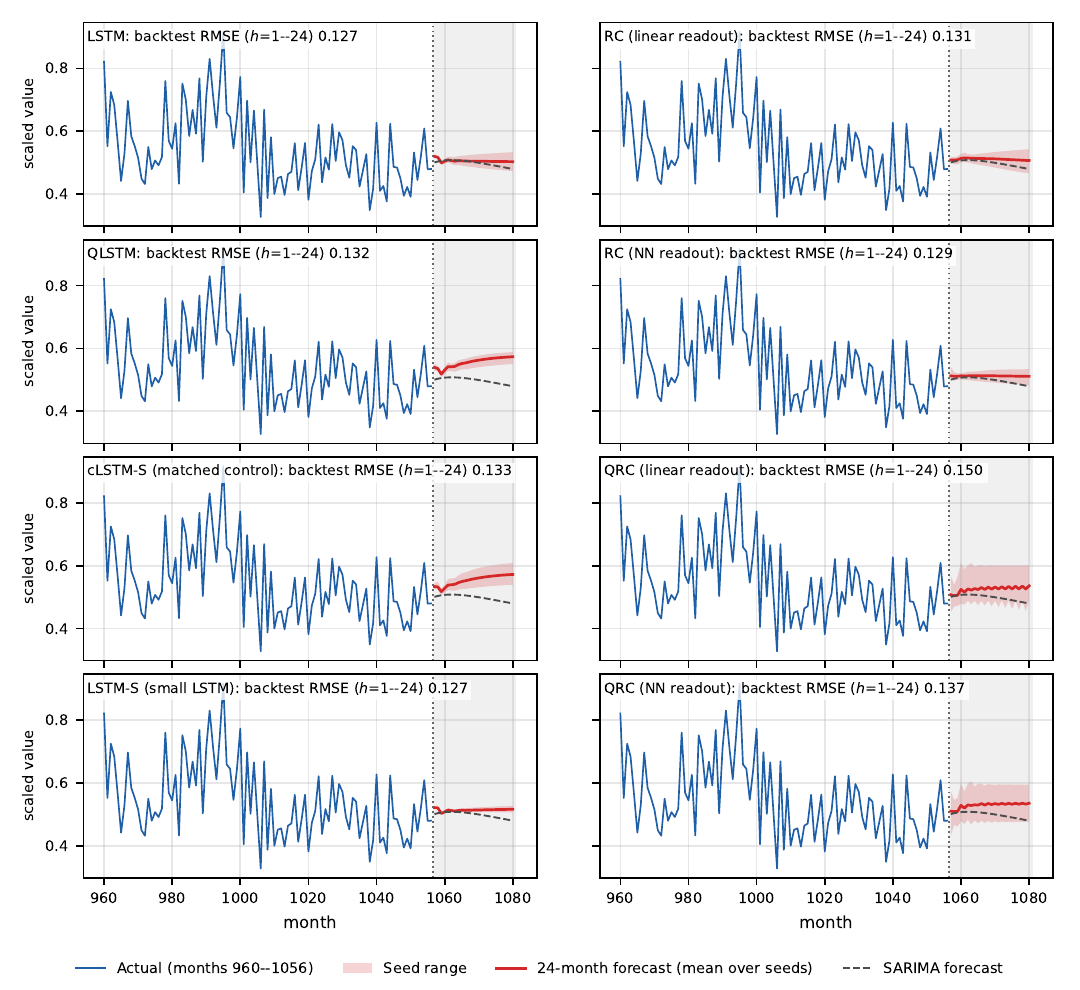}
\caption{24-month recursive forecasts of months 1057--1080 (red line: mean over 10 seeds; shaded: range over seeds) after the last 96 observed months (blue), with the SARIMA forecast (dashed). These months have no observations; the panel labels give the rolling-origin backtest RMSE over $h=1$--$24$.}
\label{fig:f24}
\end{figure}

Figure~\ref{fig:f24} shows the 24-month recursive forecasts of the unobserved months 1057--1080. The recursive forecasts converge within a few months to a smooth level and do not reproduce the month-to-month switching of the series: LSTM, LSTM-S and both RC variants stay near $0.50$--$0.52$, close to the series mean, the QLSTM and cLSTM-S drift upward to about $0.57$, the QRC forecasts settle near $0.53$ with a wide seed range, and SARIMA declines slowly to $0.48$. These forecasts illustrate the models' long-horizon behavior; they cannot be scored, and the backtest above is the evidence for their accuracy.

\begin{table}[t]
\centering
\caption{Mean wall-clock training cost per run (single CPU core, training window months 1--739). Simulated QLSTM time is not indicative of hardware cost.}
\label{tab:cost}
\tabstyle
\begin{tabular}{lrrr}
\toprule
Model & Epochs & s / epoch & Total s \\
\midrule
LSTM & 63 & 0.57 & 35.0 \\
QLSTM & 87 & 2.44 & 208.5 \\
cLSTM-S (matched control) & 83 & 0.16 & 14.1 \\
LSTM-S (small LSTM) & 137 & 0.04 & 5.0 \\
RC (linear readout) & -- & -- & 0.01 \\
RC (NN readout) & -- & -- & 0.44 \\
QRC (linear readout) & -- & -- & 0.11 \\
QRC (NN readout) & -- & -- & 0.47 \\
\bottomrule
\end{tabular}

\end{table}
Table~\ref{tab:cost} lists wall-clock training cost on one CPU core. With exact statevector simulation of four 2-qubit circuits per time step, one QLSTM epoch costs $2.44$\,s versus $0.57$\,s for the 265k-parameter LSTM and $0.16$\,s for cLSTM-S; the QLSTM's smaller parameter count does not translate into lower simulated training cost, and simulated cost says nothing about the cost on quantum hardware, where each gradient component would require repeated circuit executions. The reservoirs train in under a second.

The study uses a single series, so the seed is the only unit of replication and the results do not establish behavior across series. The validation and test windows lie entirely in the GP--HMM continuation, so the reported errors measure one-step-ahead accuracy on generator output and do not establish forecasting accuracy on observed revenue; models that match the generator's structure may be favored. The study is further limited to four lags; to exact simulation without shot noise or hardware noise; to one VQC ansatz and one classical control block; to fixed reservoir hyperparameters; and to 10 seeds. Differences of the order of $10^{-4}$ RMSE, such as those among the recurrent models, lie within run-to-run variation and are not practically meaningful, whatever their nominal significance. Finally, the underlying data are proprietary, which limits external replication to the released code and result logs.

\section{Conclusion}\label{sec:conclusion}
We compared QLSTM and QRC against a full-size LSTM, a parameter-matched classical control, a small LSTM and classical RC on a monthly revenue series extended to 1056 months by a GP--HMM generator, under a common protocol with a chronological 70/15/15 split, one-step-ahead forecasts, three baseline forecasts, ten seeds per configuration, a training-window ablation and paired statistical tests. The QLSTM, with 2,105 parameters, is statistically indistinguishable from the 265,473-parameter LSTM, but equally indistinguishable from its interface-matched classical control and from a plain LSTM of the same size. All recurrent models and the classical reservoirs improve on SARIMA by about 2\% at one step, whereas the quantum reservoir is worse than both RC and SARIMA. In 24-month recursive forecasts the learned models degrade more slowly than SARIMA, and the classical LSTMs become better than the QLSTM. We find no evidence of a quantum accuracy or parameter-efficiency advantage in this low-dimensional setting. Future work will evaluate the protocol on multiple series and on public benchmarks, and extend it to multivariate inputs, larger qubit counts, alternative ans\"atze and classical control blocks, validation-tuned reservoirs, and shot-noise and hardware-noise models.

\section*{Data Availability Statement}
The product-level revenue data analyzed in this study are proprietary and cannot be redistributed. The code and pipeline used in this study (data processing, synthetic data generation, models, statistical analysis and figure scripts) are available from the corresponding author on reasonable request. All experiments use the enumerated seeds $\{0,\dots,9\}$ for model initialization, mini-batch order, and reservoir construction.

\section*{Author Contributions}
DM: Conceptualization, Methodology, Formal analysis, Writing the manuscript.
GH: Conceptualization, Methodology of QRC, Writing (review \& editing).
MZ: Data exploring \& extending and, Writing (review \& editing).
MB: Resources, Data curation, Writing (review \& editing).
TD: Resources, Data curation, Writing (review \& editing).

\section*{Funding}
This work was supported by the German Federal Ministry of Research, Technology and Space (BMFTR) within the funding program ``Application oriented quantum computing'' under Contract No.\ 13N17159.

\section*{Conflict of Interest Statement}
MB and TD were employed by DATEV eG. The remaining authors declare that the research was conducted in the absence of any commercial or financial relationships that could be construed as a potential conflict of interest.

\section*{Generative AI Statement}
The author(s) declare that Generative AI was used in the creation of this manuscript. Claude (Anthropic; model Claude Opus 5.5) was used for language editing and polishing of the text and for LaTeX formatting of the manuscript.
It was not used to design the study, generate or analyze data, or produce the results. The authors reviewed and edited all output and take full responsibility for the content of this publication.

\bibliographystyle{plainnat}
\bibliography{references}

@article{makridakis2018,
  author  = {Makridakis, S. and Spiliotis, E. and Assimakopoulos, V.},
  title   = {Statistical and Machine Learning forecasting methods: Concerns and ways forward},
  journal = {PLoS ONE},
  volume  = {13},
  number  = {3},
  pages   = {e0194889},
  year    = {2018}
}

@article{schmitt2013,
  author  = {Schmitt, T. A. and Chetalova, D. and Sch{\"a}fer, R. and Guhr, T.},
  title   = {Non-stationarity in financial time series: Generic features and tail behavior},
  journal = {EPL},
  volume  = {103},
  number  = {5},
  pages   = {58003},
  year    = {2013}
}

@article{cont2001,
  author  = {Cont, R.},
  title   = {Empirical properties of asset returns: stylized facts and statistical issues},
  journal = {Quantitative Finance},
  volume  = {1},
  number  = {2},
  pages   = {223--236},
  year    = {2001}
}

@incollection{lux2009,
  author    = {Lux, T.},
  title     = {Stylized facts and the empirical properties of financial returns},
  booktitle = {Handbook of Financial Time Series},
  editor    = {Andersen, T. G. and others},
  publisher = {Springer},
  address   = {Berlin},
  pages     = {11--44},
  year      = {2009}
}

@article{hochreiter1997,
  author  = {Hochreiter, S. and Schmidhuber, J.},
  title   = {Long short-term memory},
  journal = {Neural Computation},
  volume  = {9},
  number  = {8},
  pages   = {1735--1780},
  year    = {1997}
}

@article{fischer2018,
  author  = {Fischer, T. and Krauss, C.},
  title   = {Deep learning with long short-term memory networks for financial market predictions},
  journal = {European Journal of Operational Research},
  volume  = {270},
  number  = {2},
  pages   = {654--669},
  year    = {2018}
}

@techreport{jaeger2001,
  author      = {Jaeger, H.},
  title       = {The ``echo state'' approach to analysing and training recurrent neural networks},
  institution = {German National Research Center for Information Technology},
  type        = {GMD Report},
  number      = {148},
  year        = {2001}
}

@article{biamonte2017,
  author  = {Biamonte, J. and Wittek, P. and Pancotti, N. and Rebentrost, P. and Wiebe, N. and Lloyd, S.},
  title   = {Quantum machine learning},
  journal = {Nature},
  volume  = {549},
  pages   = {195--202},
  year    = {2017}
}

@article{schuld2015,
  author  = {Schuld, M. and Sinayskiy, I. and Petruccione, F.},
  title   = {An introduction to quantum machine learning},
  journal = {Contemporary Physics},
  volume  = {56},
  number  = {2},
  pages   = {172--185},
  year    = {2015}
}

@book{schuld2018,
  author    = {Schuld, M. and Petruccione, F.},
  title     = {Supervised Learning with Quantum Computers},
  publisher = {Springer},
  address   = {Cham},
  year      = {2018}
}

@inproceedings{chen2022qlstm,
  author    = {Chen, S. Y.-C. and Yoo, S. and Fang, Y.-L. L.},
  title     = {Quantum long short-term memory},
  booktitle = {Proceedings of the IEEE International Conference on Acoustics, Speech and Signal Processing (ICASSP)},
  address   = {Singapore},
  pages     = {8622--8626},
  year      = {2022},
  doi       = {10.1109/ICASSP43922.2022.9747369}
}

@article{fujii2017,
  author  = {Fujii, K. and Nakajima, K.},
  title   = {Harnessing disordered-ensemble quantum dynamics for machine learning},
  journal = {Physical Review Applied},
  volume  = {8},
  pages   = {024030},
  year    = {2017}
}

@article{ghosh2019,
  author  = {Ghosh, S. and Opala, A. and Matuszewski, M. and Paterek, T. and Liew, T. C. H.},
  title   = {Quantum reservoir processing},
  journal = {npj Quantum Information},
  volume  = {5},
  pages   = {35},
  year    = {2019}
}

@article{mujal2021,
  author  = {Mujal, P. and Mart{\'i}nez-Pe{\~n}a, R. and Nokkala, J. and Garc{\'i}a-Beni, J. and Giorgi, G. L. and Soriano, M. C. and Zambrini, R.},
  title   = {Opportunities in quantum reservoir computing and extreme learning machines},
  journal = {Advanced Quantum Technologies},
  volume  = {4},
  pages   = {2100027},
  year    = {2021}
}

@article{doosti2024,
  author  = {Doosti, M. and Wallden, P. and Hamill, C. B. and Hankache, R. and Brown, O. T. and Heunen, C.},
  title   = {A brief review of quantum machine learning for financial services},
  journal = {arXiv preprint arXiv:2407.12618},
  year    = {2024}
}

@article{egger2020,
  author  = {Egger, D. J. and others},
  title   = {Quantum computing for finance: State-of-the-art and future prospects},
  journal = {IEEE Transactions on Quantum Engineering},
  volume  = {1},
  pages   = {3101724},
  year    = {2020}
}

@article{orus2019,
  author  = {Or{\'u}s, R. and Mugel, S. and Lizaso, E.},
  title   = {Quantum computing for finance: Overview and prospects},
  journal = {Reviews in Physics},
  volume  = {4},
  pages   = {100028},
  year    = {2019}
}

@article{khan2023qlstm,
  author  = {Khan, S. Z. and Muzammil, N. and Zaidi, S. M. H. and Aljohani, A. J. and Khan, H. and Ghafoor, S.},
  title   = {Quantum long short-term memory ({QLSTM}) vs. classical {LSTM} in time series forecasting: A comparative study in solar power forecasting},
  journal = {arXiv preprint arXiv:2310.17032},
  year    = {2023}
}

@article{cao2023qlstm,
  author  = {Cao, Y. and Zhou, X. and Fei, X. and Zhao, H. and Liu, W. and Zhao, J.},
  title   = {Linear-layer-enhanced quantum long short-term memory for carbon price forecasting},
  journal = {Quantum Machine Intelligence},
  volume  = {5},
  number  = {2},
  pages   = {26},
  year    = {2023}
}

@article{mahmood2024qlstm,
  author  = {Mahmood, T. and Ahmad, I. and Ansar, M. M. Z. and Darwish, J. A. and Sherwani, R. A. K.},
  title   = {Comparative study of long short-term memory ({LSTM}) and quantum long short-term memory ({QLSTM}): Prediction of stock market movement},
  journal = {arXiv preprint arXiv:2409.08297},
  year    = {2024}
}

@article{maheshwari,
  author  = {Maheshwari, D. and Hellstern, G. and Zaefferer, M. and Braun, M. and D{\"o}hler, T.},
  title   = {Learning temporal patterns in financial time series: A comparative study of quantum {LSTM} and quantum reservoir computing},
  journal = {arXiv preprint arXiv:2605.02656},
  year    = {2026}
}

@inproceedings{ceschini2022,
  author    = {Ceschini, A. and Rosato, A. and Panella, M.},
  title     = {Hybrid quantum-classical recurrent neural networks for time series prediction},
  booktitle = {Proceedings of the International Joint Conference on Neural Networks (IJCNN)},
  pages     = {1--8},
  year      = {2022}
}

@article{chen2020nurdin,
  author  = {Chen, J. and Nurdin, H. I. and Yamamoto, N.},
  title   = {Temporal information processing on noisy quantum computers},
  journal = {Physical Review Applied},
  volume  = {14},
  pages   = {024065},
  year    = {2020}
}

@article{fry2023,
  author  = {Fry, D. and Deshmukh, A. and Chen, S. Y.-C. and Rastunkov, V. and Markov, V.},
  title   = {Optimizing quantum noise-induced reservoir computing for nonlinear and chaotic time series prediction},
  journal = {Scientific Reports},
  volume  = {13},
  pages   = {19326},
  year    = {2023}
}

@article{li2026qrc,
  author  = {Li, Q. and Mukhopadhyay, C. and Bayat, A. and Habibnia, A.},
  title   = {Quantum reservoir computing for realized volatility forecasting},
  journal = {Physical Review Research},
  volume  = {8},
  number  = {2},
  pages   = {023028},
  year    = {2026}
}

@article{vitali2025,
  author  = {Vitali, G. and others},
  title   = {Quantum reservoir computing for credit card default prediction on a neutral atom platform},
  journal = {arXiv preprint arXiv:2510.04747},
  year    = {2025}
}

@article{bowles2024,
  author  = {Bowles, J. and Ahmed, S. and Schuld, M.},
  title   = {Better than classical? {T}he subtle art of benchmarking quantum machine learning models},
  journal = {arXiv preprint arXiv:2403.07059},
  year    = {2024}
}

@article{sklearn,
  author  = {Pedregosa, F. and others},
  title   = {Scikit-learn: Machine learning in {P}ython},
  journal = {Journal of Machine Learning Research},
  volume  = {12},
  pages   = {2825--2830},
  year    = {2011}
}

@book{rasmussenwilliams2006,
  author    = {Rasmussen, C. E. and Williams, C. K. I.},
  title     = {Gaussian Processes for Machine Learning},
  publisher = {MIT Press},
  address   = {Cambridge, MA},
  year      = {2006}
}

@inproceedings{gardner2018,
  author    = {Gardner, J. R. and Pleiss, G. and Bindel, D. and Weinberger, K. Q. and Wilson, A. G.},
  title     = {{GPyTorch}: Blackbox matrix-matrix {G}aussian process inference with {GPU} acceleration},
  booktitle = {Advances in Neural Information Processing Systems},
  volume    = {31},
  year      = {2018}
}

@article{rabiner1989,
  author  = {Rabiner, L. R.},
  title   = {A tutorial on hidden {M}arkov models and selected applications in speech recognition},
  journal = {Proceedings of the IEEE},
  volume  = {77},
  number  = {2},
  pages   = {257--286},
  year    = {1989},
  doi     = {10.1109/5.18626}
}

@article{baum1970,
  author  = {Baum, L. E. and Petrie, T. and Soules, G. and Weiss, N.},
  title   = {A maximization technique occurring in the statistical analysis of probabilistic functions of {M}arkov chains},
  journal = {The Annals of Mathematical Statistics},
  volume  = {41},
  number  = {1},
  pages   = {164--171},
  year    = {1970},
  doi     = {10.1214/aoms/1177697196}
}

@article{maheshwari2022,
  author  = {Maheshwari, D. and Sierra-Sosa, D. and Garcia-Zapirain, B.},
  title   = {Variational quantum classifier for binary classification: Real vs synthetic dataset},
  journal = {IEEE Access},
  volume  = {10},
  pages   = {3705--3715},
  year    = {2022},
  doi     = {10.1109/ACCESS.2021.3139323}
}

@article{mottonen2005,
  author  = {M{\"o}tt{\"o}nen, M. and Vartiainen, J. J. and Bergholm, V. and Salomaa, M. M.},
  title   = {Transformation of quantum states using uniformly controlled rotations},
  journal = {Quantum Information and Computation},
  volume  = {5},
  number  = {6},
  pages   = {467--473},
  year    = {2005}
}

@article{pennylane,
  author  = {Bergholm, V. and others},
  title   = {{PennyLane}: Automatic differentiation of hybrid quantum-classical computations},
  journal = {arXiv preprint arXiv:1811.04968},
  year    = {2018}
}

@misc{qalchemy,
  author       = {{Data Cybernetics}},
  title        = {q-alchemy-sdk-py: {P}ython {SDK} for the {Q-Alchemy} {API}},
  howpublished = {GitHub repository},
  year         = {2024}
}

@book{hyndman2021,
  author    = {Hyndman, R. J. and Athanasopoulos, G.},
  title     = {Forecasting: Principles and Practice},
  edition   = {3rd},
  publisher = {OTexts},
  address   = {Melbourne},
  year      = {2021}
}

@inproceedings{statsmodels,
  author    = {Seabold, S. and Perktold, J.},
  title     = {Statsmodels: Econometric and statistical modeling with {P}ython},
  booktitle = {Proceedings of the 9th Python in Science Conference},
  pages     = {92--96},
  year      = {2010}
}

\end{document}